\documentclass{article}

\usepackage{amsthm}
\usepackage{amsmath}
\usepackage{amssymb}
\usepackage{graphicx} 
\usepackage{wrapfig}
\usepackage[font=footnotesize]{caption}
\usepackage{subcaption}
\usepackage{enumitem}
\usepackage{graphicx}
\usepackage{multicol}
\usepackage{esint}

\newtheorem*{remark*}{Remark}

\title{Formal justification of a continuum relaxation model for one-dimensional moir\'e materials}
\author{Jingzhi (David) Zhou and Alexander B. Watson}

\usepackage[usenames,dvipsnames]{xcolor}

\newcommand{\dx}{\text{d}x}

\def\N{\mathbb{N}}

\newcommand{\de}{\ensuremath{\partial}}						
\newcommand{\dee}{\ensuremath{\textrm{d}}}
\newcommand{\pd}[2]{\ensuremath{ \frac{\de #1}{\de #2}}}

\newcommand{\inty}[4]{\ensuremath{ \int_{#1}^{#2} \! #3 \, \dee#4 }}

\begin{document}

\maketitle



\begin{abstract}
    Mechanical relaxation in moir\'e materials is often modeled by a continuum model where linear elasticity is coupled to a stacking penalty known as the Generalized Stacking Fault Energy (GSFE). We review and compute minimizers of a one-dimensional version of this model, and then show how it can be formally derived from a natural atomistic model. Specifically, we show that the continuum model emerges in the limit $\epsilon \downarrow 0$ and $\delta \downarrow 0$ while holding the ratio $\eta := \frac{\epsilon^2}{\delta}$ fixed, where $\epsilon := \frac{a}{a_\mathcal{M}}$ is the ratio of the monolayer lattice constant to the moir\'e lattice constant and $\delta := \frac{V_0}{\kappa}$ is the ratio of the typical stacking energy to the monolayer stiffness.
\end{abstract}

\section{Introduction}

The remarkable electronic properties of moir\'e materials, stackings of 2D materials with mismatched Bravais lattices, have attracted considerable attention in recent years, especially since the observation of correlated insulator and superconducting phases in ``magic angle'' twisted bilayer graphene \cite{Cao2018,Cao2018a}.

It has been proposed in, for example, \cite{Dai2016,2018CarrMassattTorrisiCazeauxLuskinKaxiras,2019YoonEngelkeCarrFangZhangCazeauxSungHovdenTsenTaniguchiWatanabeYiKimLuskinTadmorKaxirasKim,Cazeaux2020} that these properties are considerably modified by atomic relaxation, where atoms within each layer re-arrange themselves to minimize mechanical energy within the stacked structure.

In each of these works, atomic relaxation is modelled by a continuum model where linear elasticity describing the stiffness of each layer is coupled to a stacking energy penalty. This energy, known as the Generalized Stacking Fault Energy (GSFE), can be efficiently parametrized using Density Functional Theory (DFT) applied to untwisted layers. 
For concreteness, we refer to this model as the GSFE model. 

The present work has two main parts. In the first part (Sections \ref{sec:GSFE} and \ref{sec:numerics_results}), we review the GSFE model and its numerical solutions as it applies to stacked 1D chains, where the interlayer twist can be modelled as a mismatch between the layer lattice constants. We provide details of the numerical algorithm used to compute minimizers of the GSFE energy functional in Appendix \ref{sec:numerics_method}. 

In the second part (Section \ref{sec:atomistic}), we show how the GSFE model can be \emph{derived} by taking a formal atomistic-to-continuum limit from a natural atomistic energy defined through interatomic pair potentials. Such atomistic energies have been proposed as models for relaxation in moir\'e materials in, e.g., \cite{Hott2024}. We argue that the parameter regime where the GSFE model should be accurate is indeed realized in twisted bilayer graphene near to the magic angle. Note that in the present work we do not make any rigorous connection between the atomistic and continuum models, for example, by establishing convergence of their minimizers. This will be the content of future works. 


\subsection*{Acknowledgements} This work was carried out during JZ's REU project at University of Minnesota during Summer 2024 supervised by AW. JZ and AW would like to thank Michael Hott and Mitchell Luskin for stimulating discussions. AW's research was supported in part by grant NSF DMS-2406981.

\section{GSFE Model} \label{sec:GSFE}

We start by reviewing the GSFE model of atomic relaxation in one dimension following Nam and Koshino \cite{Nam2017}.

\begin{figure}[h]
    \centering
    \includegraphics[width=0.7\textwidth]{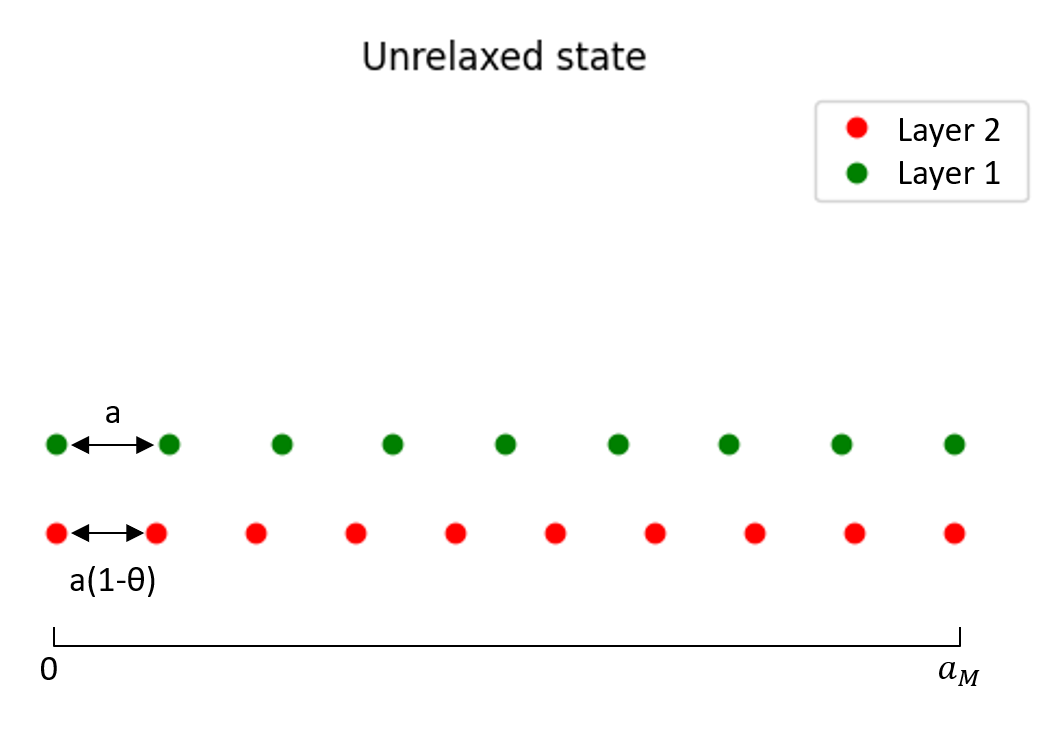}
    \caption{The two layers of atoms in one moir\'e supercell before relaxation.}
    \label{fig:unrelaxed}
\end{figure}

We consider a two layers of atoms as shown in Figure \ref{fig:unrelaxed}. In the unrelaxed state, atoms are assumed to space evenly. We assume that the lattice constant in layer 1 is $a > 0$, while the lattice constant in layer 2 is $a (1 - \theta)$, where $0 < \theta < 1$.

Starting from the origin where an atom on layer 1 is exactly aligned with an atom on layer 2, the disregistry of the $n$th atom on layer 1 relative to the $n$th atom on layer 2 is
\begin{equation} \label{eq:disreg}
    a n - a (1 - \theta) n = \theta a n \mod (1 - \theta) a.
\end{equation}
The (unrelaxed) interpolated disregistry function is obtained by replacing $an$ in \eqref{eq:disreg} by the continuous variable $x$, i.e.,
\begin{equation} \label{eq:unrelax_delta}
    \delta_0(x) := \theta x \mod (1 - \theta) a.
\end{equation}
The moir\'e period $a_\mathcal{M}$ is the minimal period of this function, so that
\begin{equation} \label{eq:moire}
    a_\mathcal{M} = \frac{a (1 - \theta)}{\theta}.
\end{equation}

We consider the case where the system is exactly periodic with respect to the moir\'e lattice. This is sometimes called making a moir\'e supercell. We thus have
\begin{equation} \label{eq:M_N_moire}
    M a = N a (1 - \theta) = a_\mathcal{M}
\end{equation}
for positive integers $M, N$. Re-arranging these identities, we have that
\begin{equation}
    N = \frac{1}{\theta},
\end{equation}
while
\begin{equation}
    M = \frac{1 - \theta}{\theta} = \frac{1}{\theta} - 1,
\end{equation}
so that $N = M + 1$.

To obtain rough physical values for these quantities, recall that the graphene lattice constant is $a = .25$ nm, while the ``magic angle'' $\theta \approx 1^\circ \approx 0.02$ rad $= \frac{1}{50}$ corresponding to $M = 49$. For these values the moir\'e cell is
\begin{equation}
    (.25) \frac{(.98)}{(.02)} \text{ nm } = 12.25 \text{ nm}.
\end{equation}
In particular, we have the separation of lengthscales $a \ll a_\mathcal{M}$. We will return to this point in Section \ref{sec:atomistic}.

We can now introduce the GSFE relaxation model following Nam and Koshino \cite{Nam2017}. Let $u_i : [0,a_\mathcal{M}) \rightarrow \mathbb{R}$, $i = 1,2$, denote moir\'e-periodic continuum atomic displacement functions, so that
\begin{equation}
    u_i(x + a_\mathcal{M}) = u_i(x), \quad i = 1,2.
\end{equation}
Note that by restricting the displacement functions to $\mathbb{R}$ we enforce that atoms move only in the parallel direction, not in the transverse direction. 

We assume that the total mechanical energy is the sum of intralayer and interlayer contributions 
\begin{equation}
    E_{\text{total}}(u_1,u_2) := E_\text{{intra}}(u_1,u_2) + E_{\text{inter}}(u_1,u_2).
\end{equation}
We model the intralayer energy by linear elasticity
\begin{equation} \label{eq:lin_elast}
    E_{\text{intra}}(u_1,u_2) := \frac{\kappa}{2} \int_0^{a_{\mathcal{M}}}\left[\left(\frac{\partial u_1}{\partial x}\right)^2+\left(\frac{\partial u_2}{\partial x}\right)^2\right]\text{d}x,
\end{equation}
where $\kappa > 0$ is an elastic constant with units of energy per unit length characterizing the stiffness of the lattices. We model the interlayer energy by
\begin{equation} \label{eq:interlayer}
    E_{\text{inter}}(u_1,u_2) := \inty{0}{a_\mathcal{M}}{ V[\delta(x)] }{x},
\end{equation}
where $V[\delta]$ is a stacking energy per unit length, and
\begin{equation}
    \delta(x) := \delta_0(x) + u_1(x) - u_2(x)
\end{equation}
denotes the interpolated disregistry function \eqref{eq:unrelax_delta} modified by relaxation. We follow Nam and Koshino by assuming the stacking energy $V$ has the simple form 
\begin{equation}
    V[\delta] = - 2 V_0 \cos\left(\frac{2 \pi}{(1 - \theta)a} \delta\right),
\end{equation}
where $V_0 > 0$. Note that assuming $V_0 > 0$ ensures that the stacking energy rewards stackings where atoms on each layer are directly on top of each other.

Again, it is useful to consider rough physical values for these quantities. According to \cite{2018CarrMassattTorrisiCazeauxLuskinKaxiras}, for graphene, we have $\kappa \approx 50,000$ meV per unit area, while values of the stacking energy $V$ vary over the scale of $20$ meV per unit area. We thus have the separation of energy scales $V_0 \ll \kappa$. We will also return to this point in Section \ref{sec:atomistic}.

Since the stacking energy depends only on the displacement difference $u_1 - u_2$, it is natural to introduce the equivalent unknown functions
\begin{equation} \label{eq:u_plus_minus}
    u_+(x) := \frac{u_1(x) + u_2(x)}{\sqrt{2}}, \quad u_-(x) := \frac{u_1(x) - u_2(x)}{\sqrt{2}},
\end{equation}
so that the energy becomes
\begin{equation}
    E_{\text{total}}(u_+,u_-) := E_{\text{intra}}(u_+,u_-) + E_{\text{inter}}(u_+,u_-),
\end{equation}
where now
\begin{equation}
    E_{\text{intra}}(u_+,u_-) := \frac{\kappa}{2} \int_0^{a_{\mathcal{M}}}\left[\left(\frac{\partial u_+}{\partial x}\right)^2+\left(\frac{\partial u_-}{\partial x}\right)^2\right]\text{d}x,
\end{equation}
and
\begin{equation}
    E_{\text{inter}}(u_+,u_-) := - 2 V_0 \inty{0}{a_\mathcal{M}}{ \cos\left(\frac{2 \pi}{(1-\theta)a} \left( \delta_0(x) + \sqrt{2} u_-(x) \right) \right) }{x}.
\end{equation}
For this functional to reach its minimum, assuming smoothness of $u_+$, we need $\frac{\partial u_+}{\partial x}=0$, which implies $u_+=\text{const}$. For simplicity, and without loss of generality, we set at this point $u_+=0$. 

After these simplifications, the minimization problem to be solved becomes finding the minimizer $u_-(x)$ for the functional 
\begin{equation} \label{eq:final_func}
    E_{\text{total}}(u_-) = \inty{0}{a_{\mathcal{M}}}{\left[\frac{\kappa}{2}\left(\frac{\partial u_-}{\partial x}\right)^2 -2V_0\cos \left(\frac{2\pi}{(1-\theta)a}\left(\delta_0(x)+\sqrt{2} u_-(x)\right)\right)\right]}{x}.
\end{equation}
For discussion of the mathematical well-posedness of this problem, see \cite{Cazeaux2020}. We computed minimizers by directly minimizing an appropriate numerical discretization of the functional. The results are shown in Section \ref{sec:numerics_results}, while the numerical method is presented in full in Appendix \ref{sec:numerics_method}. 

\section{GSFE Model: Numerical results} \label{sec:numerics_results}

In this section we discuss results of numerically computing minimizers for the GSFE functional \eqref{eq:final_func} by the method described in detail in Appendix \ref{sec:numerics_method}.

Nam and Koshino \cite{Nam2017} observed that the shape of the minimizer $u_-$ depends on the dimensionless parameter
\begin{equation} \label{eq:eta}
    \eta := \sqrt{\frac{V_0}{\kappa}}\frac{a_{\mathcal{M}}}{a}.
\end{equation}
When $\eta$ is small, we have a stiff lattice, weak interlayer interaction, or small moire period. When $\eta$ is large, we have a soft lattice, strong interlayer interaction, or large moire period. That this parameter controls the shape of minimizers comes out naturally from nondimensionalizing the model \eqref{eq:final_func} in the same way that we nondimensionalize the atomistic model we introduce in Section \ref{sec:atomistic}. In fact, there we derive \eqref{eq:EE_}, which is exactly the nondimensionalization of \eqref{eq:final_func}.

\begin{figure}[h]
\centering
    \subfloat{{\includegraphics[width=.45\textwidth]{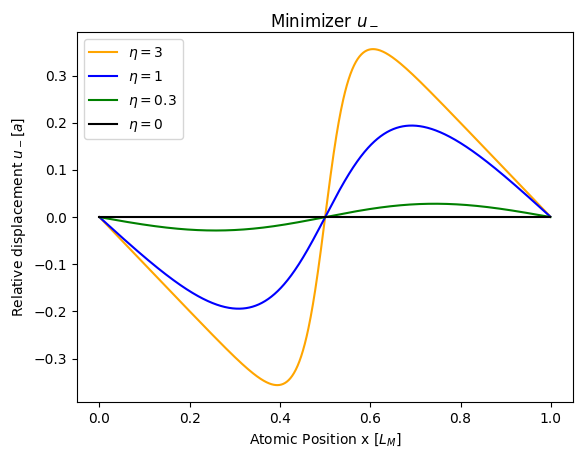} }}%
    \qquad
    \subfloat{{\includegraphics[width=.45\textwidth]{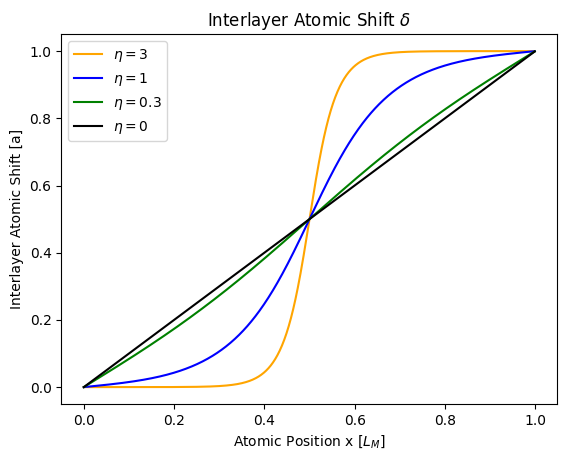} }}%
\caption{(a) Relative displacement $u_-$ and (b) interlayer atomic shift $\delta$ plotted against the position $x$ with $\eta= 3,1,0.3, \text{and } 0$.}
\label{energy}
\end{figure}

\begin{figure}[h]
\centering
    \subfloat{{\includegraphics[width=.45\textwidth]{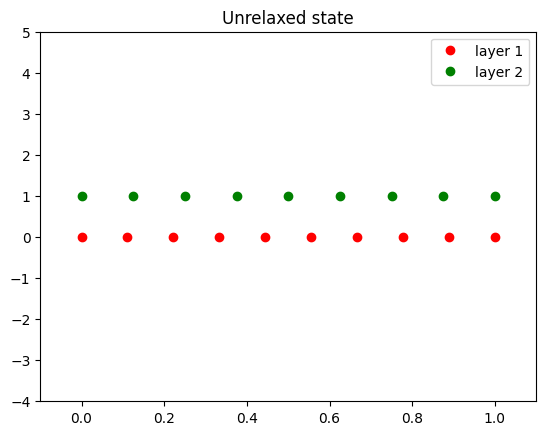} }}%
    \qquad
    \subfloat{{\includegraphics[width=.45\textwidth]{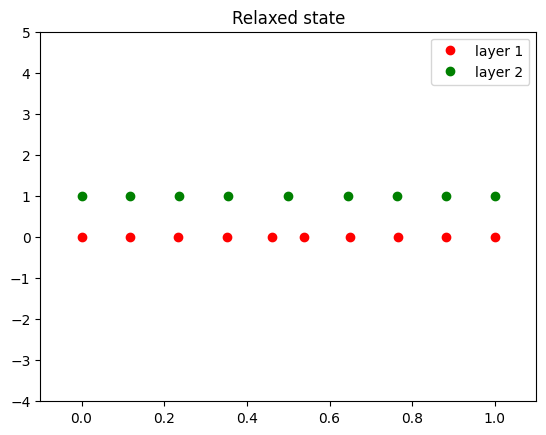} }}%
\caption{(a) Unrelaxed state (b) relaxed state with $\eta= 3$}
\label{atoms}
\end{figure}

We plot minimizers of \eqref{eq:final_func} for different values of $\eta$ in Figure \ref{energy}. We compare the unrelaxed and relaxed states of the system when $\eta = 3$ in Figure \ref{atoms}. We find results in agreement with those of Nam and Koshino \cite{Nam2017} despite them using a different method where they iteratively solved the Euler-Lagrange equation.



\section{Atomistic Model} \label{sec:atomistic}

In this section we present an alternative model of atomic relaxation based on interatomic potentials. We will show that the continuum GSFE energy functional approximates this model when $a \ll a_\mathcal{M}$ and $V_0 \ll \kappa$ while the dimensionless ratio $\eta$ (recall \eqref{eq:eta}) is held fixed.



\subsection{Atomistic energy functional}

We start by defining a natural energy functional for the bilayer system defined through interatomic potentials. For simplicity, we restrict attention to pair potentials. The approach here is influenced by \cite{Blanc2002}.

Just as before, we assume the energy has the form
\begin{equation}
    E_{\text{total}}(u_1,u_2) := E_{\text{intra}}(u_1,u_2) + E_{\text{inter}}(u_1,u_2).
\end{equation}
For $E_{\text{intra}}$, we take
 \begin{equation}
    \begin{split}
        &E_{\text{intra}}(u_1,u_2) :=  \\   
        &\frac{1}{M} \sum_{i=1}^M\sum_{\substack{j=-\infty \\  j \neq i}}^\infty W_{\text{intra}}\left(a i + u_1(a i)- a j - u_1(a j)\right)  \\
        &+ \frac{1}{M+1} \sum_{i=1}^{M+1} \sum_{\substack{j=-\infty \\  j \neq i}}^\infty W_{\text{intra}}\left(\frac{a (1-\theta)i + u_2(a (1-\theta)i)-a(1-\theta)j - u_2(a(1-\theta)j)}{1 - \theta}\right),
    \end{split}
\end{equation}
where $W_{\text{intra}}$ is an interatomic pair potential. Note that we assume periodic boundary conditions, so that we must allow for interactions between atoms within the moir\'e supercell and with their copies in all other cells. The sum over $j$ converges because of decay of $W_{\text{intra}}$. 

For $E_{\text{inter}}$, we take
\begin{equation}
    \begin{split}
        &E_{\text{inter}}(u_1,u_2) :=    \\ 
        &\frac{1}{M} \sum_{i=1}^M\sum_{j=-\infty}^\infty W_{\text{inter}}\left(a i + u_1(a i) - a (1-\theta) j - u_2(a (1-\theta) j)\right),
    \end{split}
\end{equation}
where $W_{\text{inter}}$ is another interatomic pair potential. Again, since we assume that $W_{\text{inter}}$ decays, the sum over $j$ converges. We assume at this point that $W_{\text{intra}}$ and $W_{\text{inter}}$ are both even so that
\begin{equation} \label{eq:even}
    W_{\text{intra}}(-z) = W_{\text{intra}}(z), \quad W_{\text{inter}}(-z) = W_{\text{inter}}(z).
\end{equation}

\subsection{Length non-dimensionalization}

We now nondimensionalize the model with respect to length. We introduce nondimensionalized displacement functions as
\begin{equation} \label{eq:U_nondim}
    U_i(X) := a^{-1} u_i( a_\mathcal{M} X ), \quad i = 1,2,
\end{equation}
and re-scale the interatomic potentials $W_{\text{intra}}$ and $W_{\text{inter}}$ as
\begin{equation}
    \tilde{W}_{\text{intra}}(X) := W_{\text{intra}}(a X), \quad \tilde{W}_{\text{inter}}(X) := W_{\text{inter}}(a X).
\end{equation}
With these scalings, the displacement functions $U_i$ are now $1$-periodic, and the intralayer energy becomes
 \begin{equation}
    \begin{split}
        &E_{\text{intra}}(U_1,U_2) =  \\   
        &\frac{1}{M} \sum_{i=1}^M\sum_{\substack{j=-\infty \\  j \neq i}}^\infty \tilde{W}_{\text{intra}}\left(i - j + U_1(\epsilon i) - U_1(\epsilon j)\right)  \\
        &+ \frac{1}{M+1} \sum_{i=1}^{M+1} \sum_{\substack{j=-\infty \\  j \neq i}}^\infty \tilde{W}_{\text{intra}}\left(i - j + \frac{U_2(\epsilon (1-\theta)i) - U_2(\epsilon(1-\theta)j)}{1 - \theta}\right),
    \end{split}
\end{equation}
where we introduce notation for the dimensionless length ratio 
\begin{equation} \label{eq:eps}
    \epsilon := \frac{a}{a_\mathcal{M}}.
\end{equation}
Note that \eqref{eq:moire} implies that
\begin{equation}
    \epsilon = \frac{\theta}{1 - \theta}
\end{equation}
so that when we take $\epsilon \downarrow 0$ below we must take $\theta \downarrow 0$ with $\epsilon \sim \theta$ as $\epsilon \downarrow 0$ as well. The interlayer energy becomes
\begin{equation}
    \begin{split}
        &E_{\text{inter}}(U_1,U_2) =    \\ 
        &\frac{1}{M} \sum_{i=1}^M\sum_{j=-\infty}^\infty \tilde{W}_{\text{inter}}\left(i - (1-\theta) j + U_1(\epsilon i) - U_2(\epsilon (1-\theta) j)\right).
    \end{split}
\end{equation}
We have not non-dimensionalized with respect to energy yet: both $\tilde{W}_{\text{intra}}$ and $\tilde{W}_{\text{inter}}$ still carry units of energy. We postpone this until we have simplified the model further.

\subsection{Simplification and continuum limit of the energy functional}

We now simplify the energy functional and take its continuum limit, an approximation which should be valid in the limit $\epsilon \downarrow 0$. Recalling \eqref{eq:M_N_moire}, this is equivalent to taking the limit $a \downarrow 0$ and $M \rightarrow \infty$ at fixed $a_\mathcal{M}$.

Before taking the limit, note that by re-defining $j$ and using \eqref{eq:even} the intralayer energy can be re-written as
 \begin{equation}
    \begin{split}
        &E_{\text{intra}}(U_1,U_2) =  \\   
        &\frac{1}{M} \sum_{i=1}^M\sum_{\substack{j=-\infty \\  j \neq 0}}^\infty \tilde{W}_{\text{intra}}\left(j + U_1(\epsilon i + \epsilon j) - U_1(\epsilon i)\right)  \\
        &+ \frac{1}{M+1} \sum_{i=1}^{M+1} \sum_{\substack{j=-\infty \\  j \neq 0}}^\infty \tilde{W}_{\text{intra}}\left(j + \frac{U_2(\epsilon(1-\theta)i + \epsilon(1-\theta)j) - U_2(\epsilon (1-\theta)i)}{1 - \theta}\right).
    \end{split}
\end{equation}
Taking the limit, assuming smoothness of the integrands, both sums over $i$ converge to integrals from $0$ to $1$ as
 \begin{equation}
    \begin{split}
        &\inty{0}{1}{ \sum_{\substack{j=-\infty \\  j \neq i}}^\infty \tilde{W}_{\text{intra}}\left(j + U_1(X + \epsilon j) - U_1(X)\right) }{X}  \\
        &+ \inty{0}{1}{ \sum_{\substack{j=-\infty \\  j \neq i}}^\infty \tilde{W}_{\text{intra}}\left(j + \frac{U_2(X + \epsilon(1-\theta)j) - U_2(X)}{1 - \theta}\right) }{X}.
    \end{split}
\end{equation}
We can further approximate the energy functional by Taylor-expanding the $U_i$s in powers of $\epsilon$ and keeping only the leading terms as
\begin{equation}
    \inty{0}{1}{ \tilde{W}_{\text{CB}}\left( \epsilon \de_X U_1(X) \right) + \tilde{W}_{\text{CB}}\left( \epsilon \de_X U_2(X) \right) }{X},
\end{equation}
where
\begin{equation}
    \tilde{W}_{\text{CB}}(z) := \sum_{\substack{j=-\infty \\  j \neq i}}^\infty \tilde{W}_{\text{intra}}\left(j (1 + z) \right)
\end{equation}
is the Cauchy-Born energy density. We can then further approximate by Taylor-expanding $\tilde{W}_{\text{CB}}$ in powers of $\epsilon \de_X U_i$. Assuming that constant displacement is a non-degenerate minimizer for $\tilde{W}_{\text{CB}}$, we have
\begin{equation}
    \tilde{W}_{\text{CB}}(z) = \tilde{W}_{\text{CB}}(0) + \frac{\tilde{\kappa} z^2}{2} + o(z^2) \quad z \downarrow 0,
\end{equation}
where $\tilde{\kappa} := \de_z^2 \tilde{W}_{\text{CB}}(0)$ again characterizes the stiffness of the lattices. Since its value does not affect the form of minimizers, we at this point set $\tilde{W}_{\text{CB}}(0) = 0$ without loss of generality. We arrive at the approximation
\begin{equation} \label{eq:mic_intra}
    E_{\text{intra}}(U_1,U_2) \approx \frac{\tilde{\kappa} \epsilon^2}{2} \inty{0}{1}{ \left[ \left( \pd{U_1}{X} \right)^2 + \left( \pd{U_2}{X} \right)^2 \right] }{X}.
\end{equation}

We claim that the derivation leading to \eqref{eq:mic_intra} provides a microscopic justification of the linear elasticity energy \eqref{eq:lin_elast} introduced in the GSFE model. To see this, note that upon inverting the transformation \eqref{eq:U_nondim} and setting\footnote{To see why this makes sense, note that $\kappa$ is defined as an energy per unit length, while $\tilde{\kappa}$ is defined through the derivative of an energy with respect to a nondimensionalized length and hence has units only of energy.} $\tilde{\kappa} = a_\mathcal{M} \kappa$, we obtain exactly \eqref{eq:lin_elast}.

We now consider the interlayer energy. Again, by re-defining $j$, this energy can be re-written as
\begin{equation}
    \begin{split}
        &E_{\text{inter}}(U_1,U_2) =    \\ 
        &\frac{1}{M} \sum_{i=1}^M\sum_{j=-\infty}^\infty \tilde{W}_{\text{inter}}\left(\theta i - (1-\theta) j + U_1(\epsilon i) - U_2(\epsilon (1-\theta) (i + j))\right).
    \end{split}
\end{equation}
In order to take the continuum limit we write everything in terms of the variable $\epsilon i$, so that
\begin{equation}
    \theta i = \frac{\theta}{\epsilon} \epsilon i.
\end{equation}
Recalling \eqref{eq:moire} and \eqref{eq:eps}, we have that 
\begin{equation}
    \frac{\theta}{\epsilon} \epsilon i = (1 - \theta) \epsilon i.
\end{equation}
Substituting this and taking the continuum limit we obtain
\begin{equation}
    \inty{0}{1}{ \sum_{j=-\infty}^\infty \tilde{W}_{\text{inter}}\left((1 - \theta) X - (1-\theta) j + U_1(X) - U_2((1-\theta) X + \epsilon (1-\theta) j))\right) }{X}.
\end{equation}
Just as with the intralayer energy, we expand $U_2$ in powers of $\epsilon$ and keep only the leading term. This yields
\begin{equation}
    \inty{0}{1}{ \tilde{V}\left( (1 - \theta) X + U_1(X) - U_2((1-\theta) X))\right) }{X},
\end{equation}
where we have introduced the $(1-\theta)$-periodic stacking energy 
\begin{equation}
    \tilde{V}(z) := \sum_{j=-\infty}^\infty \tilde{W}_{\text{inter}}\left(z - (1-\theta) j\right).
\end{equation}
Finally, we use smallness of $\theta$ to Taylor-expand $U_2$, to arrive at 
\begin{equation} \label{eq:mic_inter}
    \inty{0}{1}{ \tilde{V}\left( (1 - \theta) X + U_1(X) - U_2(X))\right) }{X}.
\end{equation}

We now claim that the derivation of \eqref{eq:mic_inter} above provides a microscopic justification of \eqref{eq:interlayer}. First, note that inverting the transformation \eqref{eq:U_nondim} and setting\footnote{Here we multiply by $a_\mathcal{M}$ for the same reason as when relating $\kappa$ and $\tilde{\kappa}$.}
\begin{equation}
    \tilde{V}(X) = a_\mathcal{M} V(a X),
\end{equation}
the energy \eqref{eq:mic_inter} becomes 
\begin{equation}
    a_\mathcal{M} \inty{0}{1}{ V\left( (1 - \theta) a X + u_1(X) - u_2(X))\right) }{X}.
\end{equation}
Changing variables in the integral as $x = a_\mathcal{M} X$, after recognizing \eqref{eq:moire} that $\theta = \frac{a (1 - \theta)}{a_\mathcal{M}}$, yields
\begin{equation}
    \inty{0}{a_\mathcal{M}}{ V\left( \theta x + u_1(x) - u_2(x))\right) }{x},
\end{equation}
which is exactly \eqref{eq:interlayer}. Note that $(1-\theta)$-periodicity of $\tilde{V}$ is exactly consistent with the $(1 - \theta) a$-periodicity of $V$.

We note further that, at the cost of additional error proportional to $\theta$, we can simplify \eqref{eq:mic_inter} to 
\begin{equation}
    \inty{0}{1}{ \tilde{V}\left( X + U_1(X) - U_2(X))\right) }{X},
\end{equation}
where $\tilde{V}$ is $1$-periodic. If we also simplify $\tilde{V}$ to the form of a single sinusoid as we did for the GSFE model we obtain the approximation   
\begin{equation} \label{eq:mic_inter_2}
    E_{\text{inter}}(U_1,U_2) \approx - 2 \tilde{V}_0 \inty{0}{1}{ \cos\left( 2 \pi \left( X + U_1(X) - U_2(X)) \right) \right) }{X},
\end{equation}
where $\tilde{V}_0 := a_\mathcal{M} V_0$.

\begin{remark*}
    We expect that all of the simplifications made so far can be made rigorous under general smoothness and decay assumptions on the pair potentials $W_{\text{\emph{intra}}}$ and $W_{\text{\emph{inter}}}$, with the exception of replacing $\tilde{V}$ by a single sinusoid. However, a more careful analysis may show that this simplification can also be made rigorous because of an effect where interlayer interactions are smoothed out by the interlayer distance and thus have rapidly decaying Fourier transforms. This effect is important in the derivation of the Bistritzer-MacDonald model of twisted bilayer graphene's electronic properties \cite{Watson2022}.
\end{remark*}

To summarize the results of this section, we have obtained an approximation of the atomistic energy functional as the sum of \eqref{eq:mic_intra} and \eqref{eq:mic_inter_2}. Just as we did with the GSFE model, it is natural to write the functional as a function of new unknowns
\begin{equation} \label{eq:u_plus_minus_2}
    U_+(x) := \frac{U_1(x) + U_2(x)}{\sqrt{2}}, \quad U_-(x) := \frac{U_1(x) - U_2(x)}{\sqrt{2}},
\end{equation}
and set (WLOG) $U_+ = 0$. We thus derive an approximation of the atomistic energy functional depending only on $U_-$ as
\begin{equation} \label{eq:E_func}
    E_{\text{total}}(U_-) \approx \inty{0}{1}{ \left[ \frac{\epsilon^2 \tilde{\kappa}}{2} \left( \pd{U_-}{X} \right)^2 - 2 \tilde{V}_0 \cos\left( 2 \pi \left(X + \sqrt{2} U_-(X) \right) \right) \right] }{X},
\end{equation}
which is nothing but a partially nondimensionalized version of \eqref{eq:final_func}.

\subsection{Energy non-dimensionalization}

After the simplifications made in the previous section, we have that the intralayer part of the functional \eqref{eq:E_func} depends only on the energy scale $\tilde{\kappa} = \kappa a_\mathcal{M}$, while the interlayer part depends only on the energy scale $\tilde{V}_0 = V_0 a_\mathcal{M}$. It is natural then to non-dimensionalize with respect to energy by introducing the dimensionless energy
\begin{equation} 
    \tilde{E}_{\text{total}} := \frac{E_{\text{total}}}{\tilde{V}_0}.
\end{equation}
We thus arrive at the fully nondimensionalized model 
\begin{equation} \label{eq:EE_}
    \tilde{E}_{\text{total}}(U_-) \approx \inty{0}{1}{ \left[ \frac{\epsilon^2}{2 \delta} \left( \pd{U_-}{X} \right)^2 - 2 \cos\left( 2 \pi \left(X + \sqrt{2} U_-(X) \right) \right) \right] }{X},
\end{equation}
where we have introduced notation for the dimensionless energy ratio
\begin{equation}
    \delta := \frac{\tilde{V}_0}{\tilde{\kappa}} = \frac{V_0}{\kappa}.
\end{equation}
It is clear that to have a model where the intralayer and interlayer terms balance in the limit $\epsilon \downarrow 0$, i.e., where neither term is asymptotically larger in this limit, we require that $\delta \downarrow 0$ in such a way that the dimensionless ratio 
\begin{equation}
    \eta := \frac{\sqrt{\delta}}{\epsilon} = \sqrt{ \frac{V_0}{\kappa} } \frac{a_\mathcal{M}}{a},
\end{equation}
remains fixed. The form of minimizers in the limit, moreover, depends only on the value of $\eta$ as the limit is taken. This justifies Nam and Koshino's observation \cite{Nam2017} that minimizers of the GSFE model depend only on the value of this parameter; recall \eqref{eq:eta}.

It is interesting to again consider physical values of $\epsilon$ and $\delta$ and check whether, for example, twisted bilayer graphene at the magic angle is in this regime. Recall that there we have $a \approx .25$ nm, and $\theta \approx \frac{1}{50} \approx 1.1^\circ$. We can then estimate
\begin{equation}
    \epsilon = \frac{a}{a_\mathcal{M}} \approx \frac{.25}{50 (.25)} = \frac{1}{50} = 0.02.
\end{equation}
As for the energy scales, we have $\kappa \approx 50,000$ meV per unit area, and $V_0 \approx 20$ meV per unit area \cite{2018CarrMassattTorrisiCazeauxLuskinKaxiras}. We thus have
\begin{equation}
    \delta = \frac{V_0}{\kappa} \approx \frac{20}{50,000} = \frac{1}{2500} = 0.0004,
\end{equation}
so that, remarkably,
\begin{equation}
    \eta \approx 1
\end{equation}
in twisted bilayer graphene. We conclude that, for twisted bilayer graphene near the magic angle, \eqref{eq:final_func} can indeed be expected to be a good model.

\bibliographystyle{plain}
\bibliography{ABW_library}

\appendix

\section{GSFE Model: Numerical Method} \label{sec:numerics_method}

We aim to find the minimizer $u_-(x)$ for the functional 
\[
E_\mathrm{intra}+E_\mathrm{inter} = \int_0^{a_{\mathcal{M}}}\frac{1}{2}k\left(\partial_x u_-\right)^2 -2V_0\cos \left(\frac{2\pi}{a}(\delta_0(x)+u_-(x))\right) \dx.
\]
First we discretize the region $[0,a_{\mathcal{M}}]$ by taking $n=0,1,\dots, N\in \N$ and define $\Delta x := \frac{a_{\mathcal{M}}}{N}$ and $x_{n}:=n\Delta x$. Now we can discretize $u_-(x)$ and define $u_{-n}:=u_-({x_n})$. The central difference method can be used to approximate $\frac{\partial u_-}{\partial x}$. Then the total energy can be approximated with the following Riemann sum,
\begin{equation}
\begin{split}\label{sum}
    &E_\mathrm{intra}+E_\mathrm{inter}\approx \\
    &\sum_{n=0}^{N-1}\left[\frac{1}{2}k\left(\frac{u_{-(n+1)}-u_{-(n-1)}}{2\Delta x}\right)^2-2V_0\cos \left(\frac{2\pi n}{N}+\frac{2\pi}{a}u_{-n}\right)\right]\Delta x.
\end{split}
\end{equation}
We assume that the superlattice remains periodic with moiré length $a_{\mathcal{M}}$. We impose periodic boundary conditions that $u_{-1}=u_{-(N-1)}$, $u_{-0}= u_{-N}$, and so on. Now the minimization problem becomes finding the vector $\mathbf{u_-}:=(u_{-0},u_{-1},\dots,u_{-(N-1)})^T$ that minimizes equation \eqref{sum}. We can use the limited-memory BFGS algorithm to find the minimizer $\mathbf{u_-}$. To use the Limited-memory BFGS algorithm, we need to compute the gradient of equation \eqref{sum}. The $n$th term of the gradient is
\begin{equation}
\begin{split}\label{grad}
    &\nabla(E_\mathrm{intra}+E_\mathrm{inter})_n = \\
    &\frac{k}{4\Delta x}\left(2u_{-n}-u_{-(n-2)}-u_{-(n+2)}\right)+\frac{4\pi}{a}V_0\sin\left(\frac{2\pi n}{N}+\frac{2\pi}{a}u_{-n}\right)\Delta x.
\end{split}
\end{equation}

We implemented this approach in the Python language on Google Colab using the function \textit{scipy.optimize.fmin\_l\_bfgs\_b} in the SciPy package to numerically minimize the total energy.

\end{document}